\documentclass[ aps, showpacs, showkeys, nofootinbib, floatfix, superscriptaddress]{revtex4}

\usepackage{amsfonts}

\usepackage{amssymb}
\usepackage{amsmath}
\usepackage{graphicx}
\usepackage{hyperref}

\begin{document}

\title{Cosmological constraints on the generalized holographic dark energy}

 \author{Jianbo Lu}
 \email{lvjianbo819@163.com}
 \affiliation{Department of Physics, Liaoning Normal University, Dalian 116029, P. R. China}
 \author{Yuting Wang}
  \affiliation{School of Physics and Optoelectronic Technology, Dalian University of Technology, Dalian, 116024, P. R. China}
 \author{Yabo Wu}
   \affiliation{Department of Physics, Liaoning Normal University, Dalian 116029, P. R. China}
 \author{Tianqiang Wang}
  \affiliation{Department of Physics, Liaoning Normal University, Dalian 116029, P. R. China}

\begin{abstract}  We use the Markov Chain
Monte Carlo method to investigate a global constraints on
generalized holographic (GH) dark energy with flat and non-flat
universe from the current observed data: the Union2 dataset of type
supernovae Ia (SNIa), high-redshift Gamma-Ray Bursts (GRBs), the
observational Hubble data (OHD), the cluster X-ray gas mass
fraction, the baryon acoustic oscillation (BAO), and the
 cosmic microwave background (CMB) data. The most stringent constraints on
   GH  model parameter are obtained. In addition, it is found that the equation of state for
 this generalized holographic dark energy can cross over the phantom boundary
 $w_{de}=-1$.
\end{abstract}


\maketitle

\section{$\text{Introduction}$}

{The late accelerating universe \cite{SNeCMBLSS}
 is often interpreted by introducing a new component dubbed as dark energy (DE) with negative
pressure in the standard  cosmology. And a natural candidate of DE
is positive tiny cosmological constant, though it suffers from both
the  fine tuning and cosmic coincidence problems.
 If DE is not a constant
but a time variable one, the fine tuning and cosmic coincidence
problems can be solved. So, lots of dynamical dark energy models
were investigated in the past years \cite{DEmodels}. Especially, the
energy density given by basing  the holographic principle are
studied extensively \cite{HDE}. According to the holographic
principle it is required that the total energy for a system with
size $L$ should not exceed the mass of a black hole of the same
size. The largest $L$ allowed indicates an energy density
$\rho_{\Lambda}=3c^{2}M_{p}^{2}L^{-2}$, where $c$ is a numerical
constant and $M_{p}$ is the reduced Planck Mass $M_{p}^{-2}=8\pi G$.
Applying this principle to cosmology, the UV cut-off is related to
the vacuum energy, and IR cut-off is related to the large scale of
the universe such as  Hubble horizon, future event horizon, particle
horizon, etc. And an accelerated universe can be gotten by taking
the future event horizon as an IR cut-off, with existing a causality
problem. Unfortunately, though the Hubble horizon is the most
natural cosmological length scale, non-accelerated universe can be
obtained \cite{HDE} when this
 horizon is taken as the IR cut-off. So, how to obtain an accelerated
expansion by using the Hubble horizon as the IR cut-off  is
interesting.

In addition, on the basis of  holographic principle  Ref.
\cite{Ricci} take the Ricci scalar as the IR cut-off and obtain a
new form of energy density,
$\rho_{R}=3c^{2}M_{p}^{2}(\dot{H}+2H^{2}+k/a^{2})\propto R$, dubbed
as  Ricci dark energy. For this model it avoid the causality problem
 and solve the coincidence problem \cite{Ricci}.
And in Ref. \cite{Cai-Ricci} it is found that the Ricci dark energy
has relation with the causal connection scale
$R^{-2}=Max(\dot{H}+2H^{2},-\dot{H})$ for a flat universe. Also, it
is shown that  for these two cases only taking
$R^{-2}=\dot{H}+2H^{2}$
 as the IR cut-off, the obtained model is consistent with the current cosmic
observations when the dark energy is looked as an independently
conserved  component $\dot{\rho}_{de}+3H(\rho_{de}+P_{de})=0$
\cite{Cai-Ricci}. And as indicated in Ref. \cite{GH-xu}, $H^{2}$ or
$\dot{H}$ alone can not provide an late accelerated universe  that
is consistent with the current cosmic observations. So, the
generalized holographic (or Ricci) dark energy model, i.e. a form of
their combination is investigated in Ref. \cite{GH-xu}. In this
paper we applying the current observed data to constrain the
generalized holographic (GH ) dark energy model by using the Markov
Chain Monte Carlo (MCMC) method.

\section{$\text{Basic equations for generalized holographic dark energy}$}
In a Friedmann-Robertson-Walker universe, when the Hubble horizon
and Ricci scalar are taken as the IR cut-off, the holographic dark
energy and Ricci dark energy are written as,
$\rho_{h}=3c^{2}M_{p}^{2}H^{2}$ and $\rho_{R}=3c^{2}M_{p}^{2}R$,
respectively. And in order to compare the holographic and the Ricci
dark energy, and to obtain an accelerated universe by using the
Hubble horizon as the IR cut-off, in  Ref. \cite{GH-xu} a
generalized version of holographic dark energy are constructed as,
\begin{equation}
\rho_{GH}=3c^{2}M_{p}^{2}f(\frac{R}{H^{2}})H^{2},
\end{equation}
where $f(x)$ is a function of the dimensionless variable
$x=R/H^{2}$, and  it is interesting  to write the function as
\cite{GH-xu},
\begin{equation}
f(\frac{R}{H^{2}})=1-\epsilon (1-\frac{R}{H^{2}}),
\end{equation}
where $\epsilon$ is parameter. For the generalized form of energy
density, when $\epsilon=0$ or $\epsilon=1$, it becomes holographic
or Ricci  dark energy density, respectively. Thus for this
generalized model, the dark energy density is expressed as
\begin{eqnarray}
 &\rho_{GH}&=3c^{2}M_{p}^{2}[1-\epsilon(1-\frac{R}{H^{2}})]H^{2}\nonumber\\
 &&=3c^{2}M_{p}^{2}[1-\epsilon(1-\frac{\dot{H}+2H^{2}}{H^{2}})]H^{2}\nonumber\\
 &&=3c^{2}M_{p}^{2}[1+\epsilon-\epsilon (1+z)\frac{1}{H}\frac{d H}{dz}]H^{2}.
\end{eqnarray}
And its dimensionless dark energy density is described,
\begin{eqnarray}
 &\Omega_{GH}&\equiv\frac{\rho_{GH}}{3M_{p}^{2}H^{2}}\nonumber\\
 &&=c^{2}[1+\epsilon-\epsilon (1+z)\frac{d \ln H}{dz}].
\end{eqnarray}
For the generalized holographic dark energy model, the corresponding
Friedmann equation can be written as,
\begin{equation}
H^{2}=H_{0}^{2}[\frac{2(\Omega_{0m}(1+z)^{3}+\Omega_{r}(1+z)^{4}+\Omega_{k}(1+z)^{2})}{2+c^{2}(\epsilon-2)}
+(1-\frac{2(\Omega_{0m}+\Omega_{r}+\Omega_{k})}{2+c^{2}(\epsilon-2)})(1+z)^{2-\frac{2}{c^{2}\epsilon}+\frac{2}{\epsilon}}],
\end{equation}
where   $\Omega_{0m}$, $\Omega_{r}$ and $\Omega_{k}$ respectively
denotes the current value of dimensionless  matter, photon and
curvature density, here $\Omega_{0m}$ include baryon matter
$\Omega_{b}$ and cold dark matter $\Omega_{c}$,
$\Omega_{0m}=\Omega_{b}+\Omega_{c}$. Furthermore, for the
deceleration parameter $q(z)$ and the geometrical diagnostic
quantity $Om(z)$ \cite{Om}, they can be expressed by the  Hubble
parameter as,
\begin{equation}
 q=-\frac{\ddot{a}}{aH^{2}}=-1+(1+z)\frac{1}{H}\frac{d  H}{dz},
\end{equation}
\begin{equation}
Om(z)\equiv\frac{E^{2}(z)-1}{x^{3}-1},~~x=1+z=\frac{1}{a},~~E(z)=\frac{H(z)}{H_{0}}.\label{23Om}
\end{equation}
And the equation of state (EOS) of generalized holographic dark
energy $w_{GH}$ can be derived as,
\begin{equation}
w_{GH}=-1+\frac{(1+z)}{3}\frac{1}{\rho_{GH}}\frac{d \rho_{GH}}{dz},
\end{equation}
according to the conservation equation with no interactions between
two dark components $\dot{\rho}_{GH}+3H(1+w_{GH})\rho_{GH}=0$.

\section{$\text{The current observed data and cosmological constraint methods}$}\label{constraint-method}
 In this part we introduce the cosmological constraint methods and the current observed data used in this paper.
 Concretely, it includes  557 Union2
dataset of type supernovae Ia (SNIa) \cite{557Union2}, 59
high-redshift Gamma-Ray Bursts (GRBs) data \cite{GRBs30},
observational Hubble data (OHD) \cite{OHD}, X-ray gas mass fraction
in  cluster \cite{ref:07060033}, baryon acoustic oscillation (BAO)
\cite{ref:Percival2}, and cosmic microwave background (CMB) data
\cite{7ywmap}.

\subsection{Type Ia supernovae}
For SNIa observations, we use the SNIa Union2 dataset that includes
$557$ SNIa \cite{557Union2}. Following
\cite{ref:smallomega,ref:POLARSKI}, one can obtain the corresponding
constraints by fitting the distance modulus $\mu(z)$,
\begin{equation}
\mu_{th}(z)=5\log_{10}[D_{L}(z)]+\mu_{0}.
\end{equation}
 In
this expression $D_{L}(z)=H_0 d_L(z)/c$ is the Hubble-free
luminosity distance, with $H_0$ being the Hubble constant described
by the re-normalized quantity $h$ as $H_0 =100 h~{\rm km ~s}^{-1}
{\rm Mpc}^{-1}$,
 and
\begin{eqnarray}
d_L(z)&=&\frac{c(1+z)}{\sqrt{|\Omega_k|}}sinn[\sqrt{|\Omega_k|}\int_0^z\frac{dz'}{H(z')}], \nonumber\\
 &&\mu_{0}=5log_{10}(\frac{H_{0}^{-1}}{Mpc})+25=42.38-5log_{10}h,\nonumber
\end{eqnarray}
where $sinnn(\sqrt{|\Omega_k|}x)$ respectively denotes
$\sin(\sqrt{|\Omega_k|}x)$, $\sqrt{|\Omega_k|}x$, and
$\sinh(\sqrt{|\Omega_k|}x)$ for $\Omega_k<0$, $\Omega_k=0$ and
$\Omega_k>0$.
 Additionally, the observed distance moduli $\mu_{obs}(z_i)$ of SNIa at $z_i$ is
\begin{equation}
\mu_{obs}(z_i) = m_{obs}(z_i)-M,
\end{equation}
where $m$ and $M$ are apparent magnitude and absolute magnitude of
SNIa.

For using SNIa data, theoretical model parameters  $p_s$ can be
determined by a likelihood analysis, based on the calculation of
\begin{eqnarray}
&\chi^2(p_s,M^{\prime})& \equiv \sum_{SNIa}\frac{\left\{
\mu_{obs}(z_i)-\mu_{th}(p_s,z_i)\right\}^2} {\sigma_i^2}\nonumber\\
&&=\sum_{SNIa}\frac{\left\{ 5 \log_{10}[D_L(p_s,z_i)] - m_{obs}(z_i)
+ M^{\prime} \right\}^2} {\sigma_i^2}, \ \ \ \ \label{eq:chi2}
\end{eqnarray}
where $M^{\prime}\equiv\mu_0+M$ is a nuisance parameter which
includes the absolute magnitude and the parameter $h$. The nuisance
  parameter $M^{\prime}$ can be marginalized over
analytically \cite{ref:SNchi2} as
\begin{equation}
\bar{\chi}^2(p_s) = -2 \ln \int_{-\infty}^{+\infty}\exp \left[
-\frac{1}{2} \chi^2(p_s,M^{\prime}) \right] dM^{\prime},\nonumber
\label{eq:chi2marg}
\end{equation}
resulting to
\begin{equation}
\bar{\chi}^2 =  A - \frac{B^2}{C} + \ln \left( \frac{C}{2\pi}\right)
, \label{eq:chi2mar}
\end{equation}
with
\begin{eqnarray}
&&A=\sum_{SNIa} \frac {\left\{5\log_{10}
[D_L(p_s,z_i)]-m_{obs}(z_i)\right\}^2}{\sigma_i^2},\nonumber\\
&& B=\sum_{SNIa} \frac {5
\log_{10}[D_L(p_s,z_i)]-m_{obs}(z_i)}{\sigma_i^2},\nonumber
\\
&& C=\sum_{SNIa} \frac {1}{\sigma_i^2}\nonumber.
\end{eqnarray}
Relation (\ref{eq:chi2}) has a minimum at the nuisance parameter
value $M^{\prime}=B/C$, which contains information of the values of
$h$ and $M$. Considering that the expression
\begin{equation}
\chi^2_{SNIa}(p_s)=A-(B^2/C),\label{eq:chi2SN}
\end{equation}
is only different from Eq. (\ref{eq:chi2mar}) with  a constant term
$\ln (C/2 \pi)$, it is often used in the likelihood analysis
\cite{ref:smallomega,ref:SNchi2}.

\subsection{High-redshift Gamma-Ray Bursts data}

 The GRBs data can be observed at higher redshift than
SNIa. The currently observed reshift range of GRBs is at $0.1
\lesssim z \lesssim 9$. Therefore, the GRBs data can be viewed as an
excellent complement to SNIa data and would provide more information
at high redshift. When several empirical relations of the GRBs are
proposed, these indicators have motivated the authors make use of
the GRBs as cosmological standard candles at high redshift. However,
the fact that there are not sufficient low reshift GRBs leads that
the calibration of GRB relations is dependent on the cosmological
model, namely, the circularity problem. One of methods to solve the
circularity problem is the calibration of GRB relations are
performed by the use of a sample of SNIa at low redshift in the
cosmology-independent way \cite{GRBs48}. Here, the GRBs data we used
consists of 59 GRB samples with a redshift range of $1.4 \lesssim z
\lesssim 9$ obtained in \cite{GRBs30}. These 59 GRBs are calibrated
by utilizing the newly released 557 Uion2 SNIa  and the isotropic
energy-peak spectral energy ($E_{iso}$- $E_{p,i}$) relation (i.e.
Amati relation) \cite{GRBs49}.

The $\chi^{2}_{GRBs}$ takes the same form as $\chi^{2}_{SNIa}$
\begin{equation}
\chi^{2}_{GRBs}(p_{s},\mu_{0})=\sum_{i=1}^{59}\frac{[\mu_{obs}(z_{i}
-\mu_{th}(z_{i};p_{s},\mu_{0})]^{2}}{\sigma_{i}^{2}}.\label{chi2GRBs}
\end{equation}
The same method are used to deal with the nuisance parameter
$\mu_{0}$ as shown in the description of $\chi^{2}_{SNIa}$ above.

\subsection{Observational Hubble data}

The observational Hubble data \cite{ohdzhang} are given by basing
the differential ages of the galaxies. In \cite{ref:JVS2003},
Jimenez {\it et al.} obtain  an independent estimate for Hubble
parameter, and use it to constrain the cosmological models. The
Hubble parameter as a function of redshift $z$ can be written in the
form of
\begin{equation}
H(z)=-\frac{1}{1+z}\frac{dz}{dt}.
\end{equation}
So, once $dz/dt$ is known, $H(z)$ is obtained directly. By using the
differential ages of passively-evolving galaxies, Refs.
\cite{12Hubbledata,H0prior,OHD} obtain twelve values of $H(z)$
  at different
redshift (redshift interval $0\lesssim z \lesssim 1.8$), as listed
in Table \ref{table-12Hubbledata}.
\begin{table}[ht]
\begin{center}
\begin{tabular}{c|llllllllllll}
\hline\hline
 $z$ &\ 0 & 0.1 & 0.17 & 0.27 & 0.4 & 0.48 & 0.88 & 0.9 & 1.30 & 1.43 & 1.53 & 1.75  \\ \hline
 $H(z)\ ({\rm km~s^{-1}\,Mpc^{-1})}$ &\ 74.2 & 69 & 83 & 77 & 95 & 97 & 90 & 117 & 168 & 177 & 140 & 202  \\ \hline
 $1 \sigma$ uncertainty &\ $\pm 3.6$ & $\pm 12$ & $\pm 8$ & $\pm 14$ & $\pm 17$ & $\pm 60$ & $\pm 40$
 & $\pm 23$ & $\pm 17$ & $\pm 18$ & $\pm 14$ & $\pm 40$ \\
\hline\hline
\end{tabular}
\end{center}
\caption{\label{table-12Hubbledata} The observational $H(z)$
data~\cite{12Hubbledata,H0prior}.}
\end{table}
In addition, in \cite{3Hubbledata} the authors take the BAO scale as
a standard ruler in the radial direction, and obtain three more
additional data: $H(z=0.24)=79.69\pm2.32, H(z=0.34)=83.8\pm2.96,$
and $H(z=0.43)=86.45\pm3.27$.

 The values of   model parameters  can be determined according to the observational Hubble data by minimizing \cite{chi2hub}
 \begin{equation}
 \chi_{OHD}^2(H_{0},p_{s})=\sum_{i=1}^{15} \frac{[H_{th}(H_{0},p_{s};z_i)-H_{obs}(z_i)]^2}{\sigma^2(z_i)},\label{chi2OHD}
 \end{equation}
 where  $H_{th}$ is the predicted value of the Hubble parameter, $H_{obs}$ is the observed value, $\sigma(z_i)$ is the standard
 deviation measurement uncertainty, and the summation is over the $15$ observational Hubble data points at redshifts $z_i$.

\subsection{The X-ray gas mass fraction}
The X-ray gas mass fraction, $f_{gas}$, is defined as the ratio of
the X-ray gas mass to the total mass of a cluster, which is
approximately independent on the redshift for the hot ($kT \gtrsim
5keV$), dynamically relaxed clusters at the radii larger than the
innermost core $r_{2500}$. As investigated  in \cite{ref:07060033},
the $\Lambda$CDM model is much favored and is chosen as the
referenced cosmology. The model fitted to the referenced
$\Lambda$CDM data is presented as \cite{ref:07060033}
\begin{eqnarray}
&&f_{gas}^{\Lambda CDM}(z)=\frac{K A \gamma
b(z)}{1+s(z)}\left(\frac{\Omega_b}{\Omega_{0m}}\right)
\left[\frac{D_A^{\Lambda CDM}(z)}{D_A(z)}\right]^{1.5},\ \ \ \
\label{eq:fLCDM}
\end{eqnarray}
where $D_{A}^{\Lambda CDM} (z)$ and $D_{A}(z)$ denote respectively
the proper angular diameter distance in the $\Lambda$CDM  cosmology
and the current constraint model. $A$ is the angular correction
factor, which is caused by the change in angle for the current test
model $\theta_{2500}$ in comparison with that of the reference
cosmology $\theta_{2500}^{\Lambda CDM}$:
\begin{eqnarray}
&&A=\left(\frac{\theta_{2500}^{\Lambda
CDM}}{\theta_{2500}}\right)^\eta \approx
\left(\frac{H(z)D_A(z)}{[H(z)D_A(z)]^{\Lambda CDM}}\right)^\eta,
\end{eqnarray}
here, the index $\eta$ is the slope of the $f_{gas}(r/r_{2500})$
data within the radius $r_{2500}$, with the best-fit average value
$\eta=0.214\pm0.022$ \cite{ref:07060033}. And the proper (not
comoving) angular diameter distance is given by
\begin{eqnarray}
&&D_A(z)=\frac{c}{(1+z)\sqrt{|\Omega_k|}}\mathrm{sinn}[\sqrt{|\Omega_k|}\int_0^z\frac{dz'}{H(z')}].
\end{eqnarray}
It is clear that this quantity is related with $d_{L}(z)$ by
\begin{equation}
D_A(z)=\frac{d_{L}(z)}{(1+z)^2}.\nonumber
\end{equation}
In equation (\ref{eq:fLCDM}), the parameter $\gamma$ denotes
permissible departures from the assumption of hydrostatic
equilibrium, due to non-thermal pressure support; the bias factor
$b(z)= b_0(1+\alpha_b z)$ accounts for uncertainties in the cluster
depletion factor; $s(z)=s_0(1 +\alpha_s z)$ accounts for
uncertainties of the baryonic mass fraction in stars and a Gaussian
prior for $s_0$ is employed, with $s_0=(0.16\pm0.05)h_{70}^{0.5}$
\cite{ref:07060033}; the factor $K$ is used to describe the combined
effects of the residual uncertainties, such as the instrumental
calibration and certain X-ray modelling issues, and a Gaussian prior
for the 'calibration' factor is considered by $K=1.0\pm0.1$
\cite{ref:07060033}.

Following the method in Ref. \cite{ref:CBFchi21,ref:07060033} and
adopting the updated 42 observational $f_{gas}$ data in Ref.
\cite{ref:07060033}, the   values of model parameters for the X-ray
gas mass fraction analysis are determined by minimizing,
\begin{eqnarray}
&&\chi^2_{CBF}=\sum_i^N\frac{[f_{gas}^{\Lambda
CDM}(z_i)-f_{gas}(z_i)]^2}{\sigma_{f_{gas}}^2(z_i)}+\frac{(s_{0}-0.16)^{2}}{0.0016^{2}}
+\frac{(K-1.0)^{2}}{0.01^{2}}+\frac{(\eta-0.214)^{2}}{0.022^{2}},\label{eq:chi2fgas}
\end{eqnarray}
where $\sigma_{f_{gas}}(z_{i})$ is the statistical uncertainties
(Table 3 of \cite{ref:07060033}). As pointed out in
\cite{ref:07060033}, the acquiescent systematic uncertainties have
been considered according to the parameters i.e. $\eta$, $b(z)$,
$s(z)$ and $K$.

\subsection{Baryon acoustic oscillation}

The baryon acoustic oscillations are detected in the clustering of
the combined 2dFGRS  and   SDSS main galaxy samples, which measure
the distance-redshift relation at $z_{BAO} = 0.2$ and $z_{BAO} =
0.35$. The observed scale of the BAO calculated from these samples,
are analyzed using estimates of the correlated errors to constrain
the form of the distance measure $D_V(z)$
\cite{ref:Okumura2007,ref:Percival2}
\begin{equation}
 D_V(z)=[(1+z)^2 D^{2}_{A}(z) \frac{cz}{H(z;p_{s})}]^{1/3}
             =H_{0}[\frac{z}{E(z;p_{s})}(\int ^{z}_{0}\frac{dz^{'}}{E(z^{'};p_{s})})^{2}]^{\frac{1}{3}}.\label{eq:DV}
\end{equation}
In this expression $E(z;p_{s})=H(z;p_{s})/H_{0}$. The peak positions
of the BAO depend on the ratio of $D_V(z)$ to the sound horizon size
at the drag epoch (where baryons were released from photons) $z_d$,
which can be obtained by using a fitting formula
\begin{eqnarray}
&&z_d=\frac{1291(\Omega_{0m}h^2)^{-0.419}}{1+0.659(\Omega_{0m}h^2)^{0.828}}[1+b_1(\Omega_bh^2)^{b_2}],
\end{eqnarray}
with
\begin{eqnarray}
&&b_1=0.313(\Omega_{0m}h^2)^{-0.419}[1+0.607(\Omega_{0m}h^2)^{0.674}], \\
&&b_2=0.238(\Omega_{0m}h^2)^{0.223}.
\end{eqnarray}
In this paper, we use the data of $r_s(z_d)/D_V(z)$ extracted from
the Sloan Digitial Sky Survey (SDSS) and the Two Degree Field Galaxy
Redshift Survey (2dFGRS) \cite{ref:Okumura2007}, which are listed in
Table \ref{baodata}, where $r_s(z)$ is the comoving sound horizon
size
\begin{eqnarray}
r_s(z)&&{=}c\int_0^t\frac{c_sdt}{a}=c\int_0^a\frac{c_sda}{a^2H}=c\int_z^\infty
dz\frac{c_s}{H(z)} \nonumber\\
&&{=}\frac{c}{\sqrt{3}}\int_{0}^{1/(1+z)}\frac{da}{a^2H(a)\sqrt{1+(3\Omega_b/(4\Omega_\gamma)a)}},
\end{eqnarray}
where $c_s$ is the sound speed of the photon$-$baryon fluid
\begin{eqnarray}
&&c_s^{-2}=3+\frac{4}{3}\times\frac{\rho_b(z)}{\rho_\gamma(z)}=3+\frac{4}{3}\times(\frac{\Omega_b}{\Omega_\gamma})a.
\end{eqnarray}

\begin{table}[htbp]
\begin{center}
\begin{tabular}{c|l}
\hline\hline
 $z$ &\ $r_s(z_d)/D_V(z)$  \\ \hline
 $0.2$ &\ $0.1905\pm0.0061$  \\ \hline
 $0.35$  &\ $0.1097\pm0.0036$  \\
\hline
\end{tabular}
\end{center}
\caption{\label{baodata} The observational $r_s(z_d)/D_V(z)$
data~\cite{ref:Percival2}.}
\end{table}
Using the data of BAO in Table \ref{baodata} and the inverse
covariance matrix $V^{-1}$ in \cite{ref:Percival2}:
\begin{eqnarray}
&&V^{-1}= \left(
\begin{array}{cc}
 30124.1 & -17226.9 \\
 -17226.9 & 86976.6
\end{array}
\right),
\end{eqnarray}
  the $\chi^2_{BAO}(p_s)$ is given as
\begin{equation}
\chi^2_{BAO}(p_s)=X^tV^{-1}X,\label{chi2-BAO}
\end{equation}
where $X$ is a column vector formed from the values of theory minus
the corresponding observational data, with
\begin{eqnarray}
&&X= \left(
\begin{array}{c}
 \frac{r_s(z_d)}{D_V(0.2)}-0.1905 \\
 \frac{r_s(z_d)}{D_V(0.35)}-0.1097
\end{array}
\right),
\end{eqnarray}
and $X^t$ denotes its transpose.

\subsection{Cosmic microwave background}

The CMB shift parameter $R$ is provided by
\begin{equation}
 R=\sqrt{\Omega_{0m} H^2_0}(1+z_{\ast})D_A(z_{\ast})/c=\sqrt{\Omega_{0m}}\int_{0}^{z_{\ast}}\frac{H_{0}dz^{'}}{H(z^{'};p_{s})},\label{R-CMB}
\end{equation}
here, the redshift $z_{\ast}$ (the decoupling epoch of photons) is
obtained using the fitting function
\begin{equation}
z_{\ast}=1048\left[1+0.00124(\Omega_bh^2)^{-0.738}\right]\left[1+g_1(\Omega_{0m}
h^2)^{g_2}\right],\nonumber
\end{equation}
where the functions $g_1$ and $g_2$ read
\begin{eqnarray}
g_1&=&0.0783(\Omega_bh^2)^{-0.238}\left(1+ 39.5(\Omega_bh^2)^{0.763}\right)^{-1},\nonumber \\
g_2&=&0.560\left(1+ 21.1(\Omega_bh^2)^{1.81}\right)^{-1}.\nonumber
\end{eqnarray}
In addition, the acoustic scale is related to a distance ratio,
$D_A(z)/r_s(z)$,   and at decoupling epoch it is defined as
\begin{eqnarray}
&&l_A\equiv(1+z_{\ast})\frac{\pi
D_A(z_{\ast})}{r_s(z_{\ast})},\label{la}
\end{eqnarray}
where Eq.(\ref{la}) arises a factor  $1+z_{\ast}$, because $D_A(z)$
is the proper angular diameter distance, whereas $r_{s}(z_{\ast})$
is the comoving sound horizon. Using the data of $l_A, R, z_\ast$ in
\cite{7ywmap} and their covariance matrix of $[l_A(z_\ast),
R(z_\ast), z_\ast]$ (please see table \ref{tab:7yearWMAPdata} and
\ref{tab:7yearWMAPcovariance}), we can calculate the likelihood $L$
as $\chi^2_{CMB}=-2\ln L$:
\begin{eqnarray}
&&\chi^2_{CMB}=\bigtriangleup d_i[Cov^{-1}(d_i,d_j)[\bigtriangleup
d_i]^t],\label{chi2-CMB}
\end{eqnarray}
where $\bigtriangleup d_i=d_i-d_i^{data}$ is a row vector, and
$d_i=(l_A, R, z_\ast)$.\\

\begin{figure}[ht]
  \includegraphics[width=20cm]{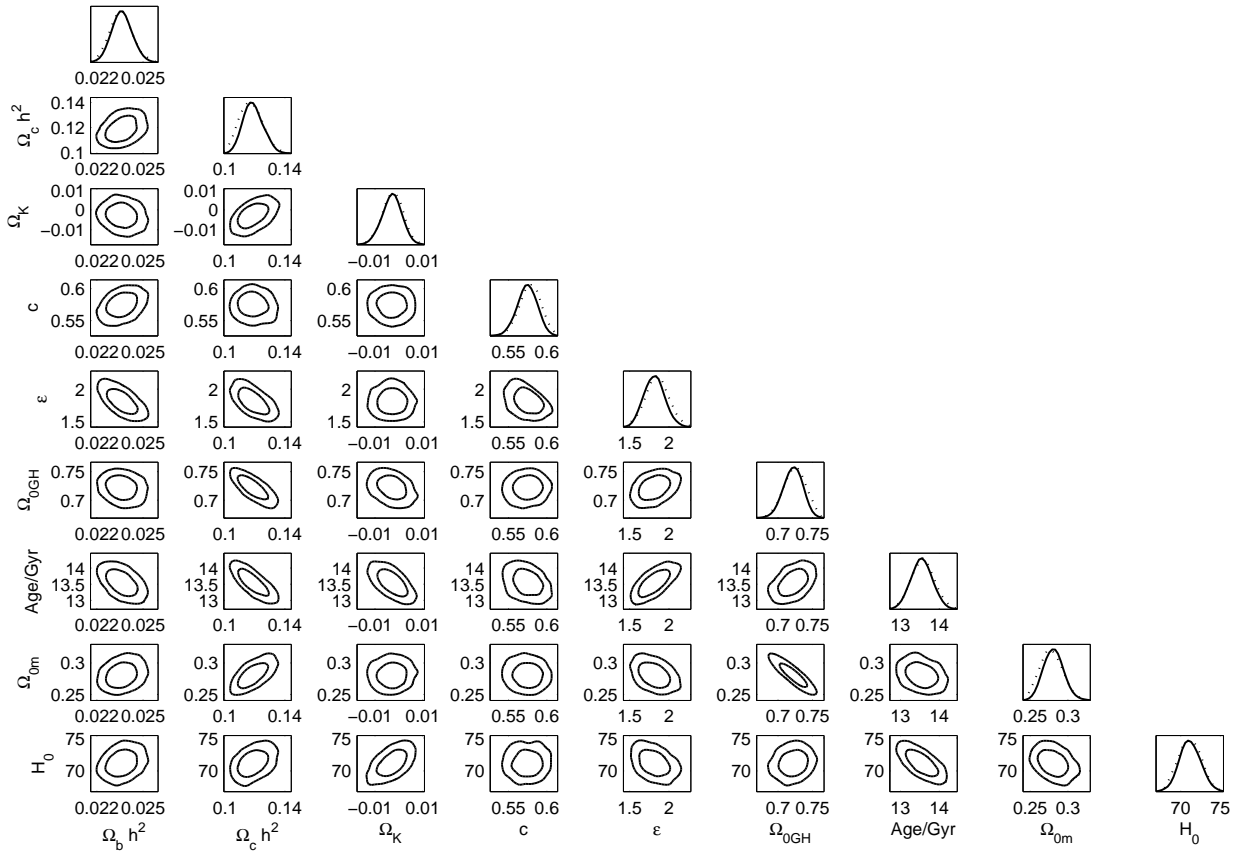}
  \caption{The 2-D contours with  $1\sigma, 2\sigma$ confidence levels and 1-D
  distribution of model parameters  in the non-flat GH model.
  Solid lines are mean likelihoods
  of samples, and dotted  lines are marginalized probabilities for 1D
  distribution.}\label{figure-ab-GH-mcmc-6data-non-flat}
\end{figure}

 \begin{table}
 \begin{center}
 \begin{tabular}{c c   cc   } \hline\hline
 ~ &              7-year maximum likelihood ~~~ & error, $\sigma$ &\\ \hline
 $ l_{A}(z_{\ast})$         & 302.09      & 0.76  & \\
 $ R(z_{\ast})$             &  1.725      & 0.018 & \\
 $ z_{\ast}$                & 1091.3     & 0.91&    \\
 \hline\hline
 \end{tabular}
 \caption{The values of  $ l_{A}(z_{\ast})$, $R(z_{\ast})$, and $z_{\ast}$ from 7-year WMAP data.}\label{tab:7yearWMAPdata}
 \end{center}
 \end{table}

\begin{table}
 \begin{center}
 \begin{tabular}{c c   cc c  } \hline\hline
 ~ &             $ l_{A}(z_{\ast})$      & $ R(z_{\ast})$   & $ z_{\ast}$ &  \\ \hline
 $ l_{A}(z_{\ast})$         & 2.305      & 29.698           &  -1.333     & \\
 $ R(z_{\ast})$             &  ~         & 6825.270         &  -113.180    & \\
 $ z_{\ast}$                & ~          & ~                &  3.414      &  \\
 \hline\hline
 \end{tabular}
 \caption{The inverse covariance matrix of  $ l_{A}(z_{\ast})$, $R(z_{\ast})$, and $z_{\ast}$ from 7-year WMAP data.}\label{tab:7yearWMAPcovariance}
 \end{center}
 \end{table}

\section{$\text{Observed constraints on generalized holographic  DE model by
using MCMC method}$}

\begin{figure}[ht]
  \includegraphics[width=18cm]{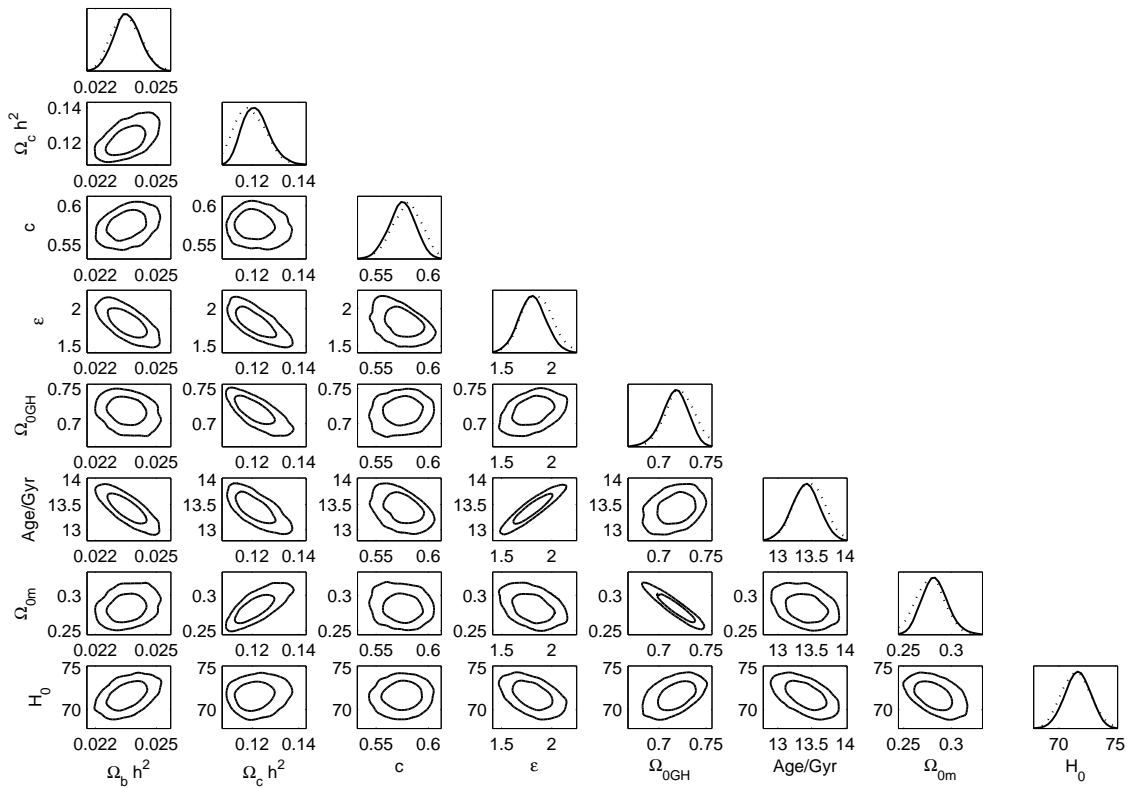}
  \caption{The 2-D contours with  $1\sigma, 2\sigma$ confidence levels and 1-D
  distribution of model parameters  in the flat GH model.
  Solid lines are mean likelihoods
  of samples, and dotted  lines are marginalized probabilities for 1D
  distribution.}\label{figure-ab-GH-mcmc-6data-flat}
\end{figure}

Next we apply the Markov Chain Monte Carlo  method to investigate a
global constraint on above generalized holographic dark energy
model. The MCMC source code can be found in the CosmoMC package
\cite{ref:MCMC} and the modified CosmoMC package
\cite{ref:0409574,ref:07060033,ref:modifiedMCMC} (this package is
about the constraint code  of X-ray cluster gas mass fraction).  To
get the converged results,  in
 MCMC calculation   we test the convergence of the chains by
taking $R - 1$ to be less than 0.03.
 The total $\chi^{2}$ is expressed as,
 \begin{equation}
 \chi^{2}_{total}(p_{s})=\chi^{2}_{SNIa}+\chi^{2}_{GRBs}+\chi^{2}_{OHD}+\chi^{2}_{CBF}+\chi^{2}_{BAO}+\chi^{2}_{CMB},\label{chi2total-4dynamical}
 \end{equation}
 with the parameter vector reading
\begin{equation}
 p_s=\{\Omega_{b}h^2, \Omega_{c}h^2,\Omega_{k}, \epsilon,c\}.
 \end{equation}
Here the expression of  $\chi^{2}$ for each observation corresponds
to Eqs.(\ref{eq:chi2SN}), (\ref{chi2GRBs}),
 (\ref{chi2OHD}), (\ref{eq:chi2fgas}), (\ref{chi2-BAO}) and (\ref{chi2-CMB}).
 Based on the  basic cosmological parameters $p_{s}$ we can  also obtain the
derived parameters $\Omega_{0m}=\Omega_{b}+\Omega_{c}$,
$\Omega_{0GH}=1-\Omega_{0m}-\Omega_{k}$, and the Hubble constant
$H_0=100h$ km$\cdot$s $^{-1}\cdot$Mpc$^{-1}$.
 Using the currently observed data with the $\chi^{2}_{total }$ in Eq. (\ref{chi2total-4dynamical}),
  Figs. \ref{figure-ab-GH-mcmc-6data-non-flat} and \ref{figure-ab-GH-mcmc-6data-flat} plot
   the 2-D contours with  $1\sigma, 2\sigma$
  confidence levels and 1-D
  distribution of model parameters  in the flat and non-flat generalized
  holographic dark energy model.
  Solid lines are mean likelihoods
 of samples, and dotted  lines are marginalized probabilities for 1D
 distribution.
  Table \ref{tableGH-mcmc-values} lists the MCMC calculation results for the constraint on model parameters.
  It includes the means, standard deviations
   with the marginalized limits for the model parameters, and  the  values for the best-fit
 sample, and projections of the n-dimensional $1\sigma$ and $2\sigma$
 confidence regions.
 The n-D limits
 give  some idea of the range of the posterior, and are much more
 conservative than the marginalized limits \cite{ref:MCMC}. From the table
 \ref{tableGH-mcmc-values} it can be seen that for the non-flat
 universe, the best fit results are given as
 $\Omega_{k}=-0.0047^{+0.0132+0.0159}_{-0.0089-0.0120}$,
 $c=0.576^{+0.034+0.037}_{-0.036-0.053}$,
 $\epsilon=1.849^{+0.347+0.461}_{-0.380-0.442}$,
 $\Omega_{0m}=0.280^{+0.036+0.047}_{-0.032-0.040}$
 (it has a smaller value of $\Omega_{0m}$ relative to the case of
 constraints on the Ricci dark energy model \cite{C-Ricci,C-Ricci1}),  with $\chi_{min}^{2}=619.314$.
  And for this case, it predicts
 the age of universe $t_{age}=13.711^{+0.709+0.924}_{-0.859-0.978}$(Gyr). Furthermore, comparing
  the Ref. \cite{GH-xu} one can see that for the generalized holographic dark energy
 the more stringent constraint on  model
 parameters at 2$\sigma$
 confidence level are
 given in this paper by using the more observational data, and it tends to have a smaller value  of
  dimensionless matter density $\Omega_{0m}$ and a bigger value of  model parameter $\epsilon$.

\begin{table}[ht]
 \vspace*{-12pt}
 \begin{center}
 \begin{tabular}{c ||  c|  c ||  c|  c} \hline\hline
 ~&        Non-flat & Non-flat  & Flat  &    Flat \\\hline\hline
 Parameters  & Best fit values &  Means & Best fit values &  Means     \\\hline
 $\Omega_{b}h^{2}$  &$0.0233^{+0.0023+0.0027}_{-0.0016-0.0016}$  & $0.0236^{+0.0006+0.0013}_{-0.0006-0.0012}$
 &$0.0236^{+0.0018+0.0022}_{-0.0017-0.0022}$  & $0.0236^{+0.0006+0.0012}_{-0.0007-0.0012}$ \\\hline
 $\Omega_{c}h^{2}$  &$0.1150^{+0.0220+0.0290}_{-0.0160-0.0160}$  & $0.1188^{+0.0067+0.0133}_{-0.0065-0.0120}$
 &$0.1178^{+0.0195+0.0241}_{-0.0105-0.0121}$  & $0.1217^{+0.0055+0.0126}_{-0.0056-0.0097}$\\\hline
 $\Omega_{k}$   &$-0.0047^{+0.0132+0.0159}_{-0.0089-0.0120}$  & $-0.0029^{+0.0040+0.0077}_{-0.0040-0.0168}$
 &----&----\\\hline
 $c$          &$0.576^{+0.034+0.037}_{-0.036-0.053}$   & $0.574^{+0.013+0.024}_{-0.012-0.026}$
 &$0.586^{+0.023+0.027}_{-0.043-0.052}$   & $0.575^{+0.012+0.023}_{-0.013-0.025}$ \\\hline
 $\epsilon$   &$1.849^{+0.347+0.461}_{-0.380-0.442}$   & $1.815^{+0.126+0.262}_{-0.129-0.237}$
 &$1.843^{+0.392+0.461}_{-0.347-0.429}$   & $1.808^{+0.130+0.262}_{-0.129-0.255}$ \\\hline
 $\Omega_{0m}$ & $0.280^{+0.036+0.047}_{-0.032-0.040}$  & $0.281^{+0.013+0.026}_{-0.013-0.024}$
 & $0.279^{+0.034+0.051}_{-0.035-0.037}$  & $0.283^{+0.013+0.029}_{-0.013-0.024}$ \\\hline
 $\Omega_{0GH}$ & $0.725^{+0.035+0.041}_{-0.044-0.055}$  & $0.722^{+0.013+0.025}_{-0.014-0.028}$
 & $0.721^{+0.035+0.037}_{-0.034-0.051}$  & $0.717^{+0.013+0.024}_{-0.013-0.029}$\\\hline
 $H_{0}$       &$70.361^{+4.710+5.143}_{-2.611-3.651}$  &  $71.181^{+1.313+2.627}_{-1.274-2.490}$
 &$71.170^{+3.514+3.939}_{-2.369-3.347}$  &  $71.656^{+1.102+2.056}_{-1.095-2.151}$\\\hline\hline
 $t_{age}$(Gyr) &$13.711^{+0.709+0.924}_{-0.859-0.978}$  &  $13.549^{+0.263+0.524}_{-0.267-0.515}$
 &$13.462^{+0.565+0.584}_{-0.589-0.747}$  &  $13.413^{+0.198+0.381}_{-0.200-0.402}$\\\hline\hline
 \end{tabular}
 \end{center}
 \caption{ For the flat and non-flat universe, the best fit model parameters with their
 limits from the extremal values of the n-dimensional distribution (recommended);
  and the means
   with the marginalized limits for the model parameters,  from MCMC calculation by using
  SNIa Union2, GRBs, OHD, CBF, BAO, and CMB  data.  }\label{tableGH-mcmc-values}
 \end{table}

\begin{table}[ht]
 \vspace*{-12pt}
 \begin{center}
 \begin{tabular}{c |c |  c|  c |  c} \hline\hline
   &$z_{T}$ & $q_{0}$  &  $Om_{0}$  &  $w_{0GH}$   \\\hline
  Non-flat &$0.706^{+0.039}_{-0.036}$ &  $-0.639^{+0.042}_{-0.047}$  & $0.241^{+0.047}_{-0.047}$
   & $-1.051^{+0.048}_{-0.048}$ \\\hline
  Flat  & $0.705^{+0.038}_{-0.034}$ &  $-0.598^{+0.041}_{-0.042}$  & $0.268^{+0.043}_{-0.043}$
   & $-1.015^{+0.045}_{-0.045}$ \\\hline\hline
 \end{tabular}
 \end{center}
 \caption{ The best fit values of transition redshift, current values of
 deceleration parameter, $Om$ parameter, and EOS of generalized holographic dark energy with their confidence levels
 for flat and non-flat universe. }\label{quantities-results}
 \end{table}

\begin{table}[ht]
 \vspace*{-12pt}
 \begin{center}
 \begin{tabular}{c |  c|  c |  c|  c |  c|  c |  c} \hline\hline
 &  $\eta$ & $\gamma$  &  $K$  &  $b_{0}$  & $\alpha_{b}$ & $s_{0}$  & $\alpha_{s}$ \\\hline
  Non-flat  &  0.212  & 1.081 &  0.998  & 0.732  & -0.092 &  0.174  & -0.055    \\\hline
  Flat &  0.208  & 1.025 &  0.958  & 0.783  & -0.086 &  0.156  & 0.020   \\\hline\hline
 \end{tabular}
 \end{center}
 \caption{ The best fit values  of parameters in $f_{gas}$ analysis method for flat and non-flat universe. }\label{7fgas-results}
 \end{table}


\begin{figure}[ht]
  ~~ \includegraphics[width=4cm]{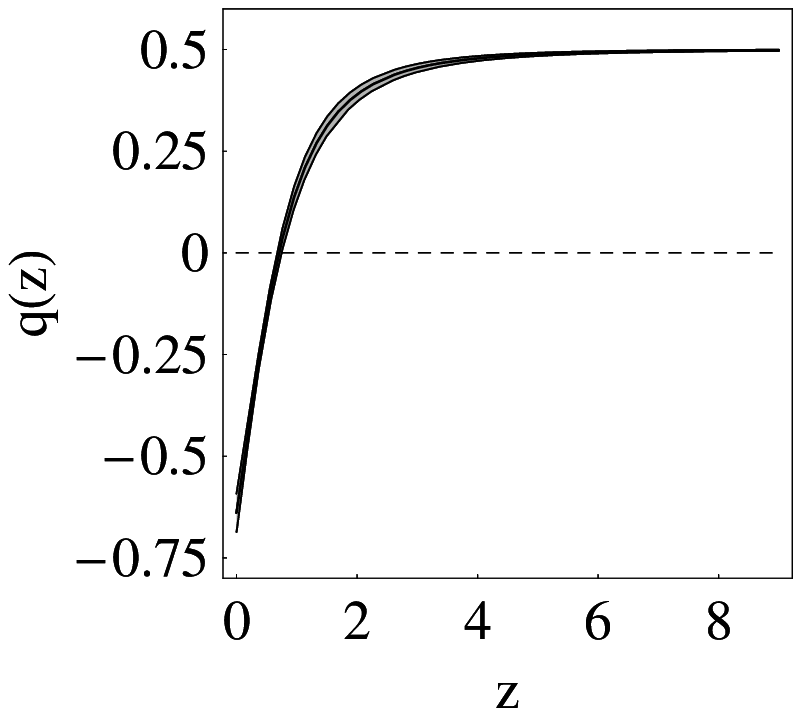}
    ~~ \includegraphics[width=4cm]{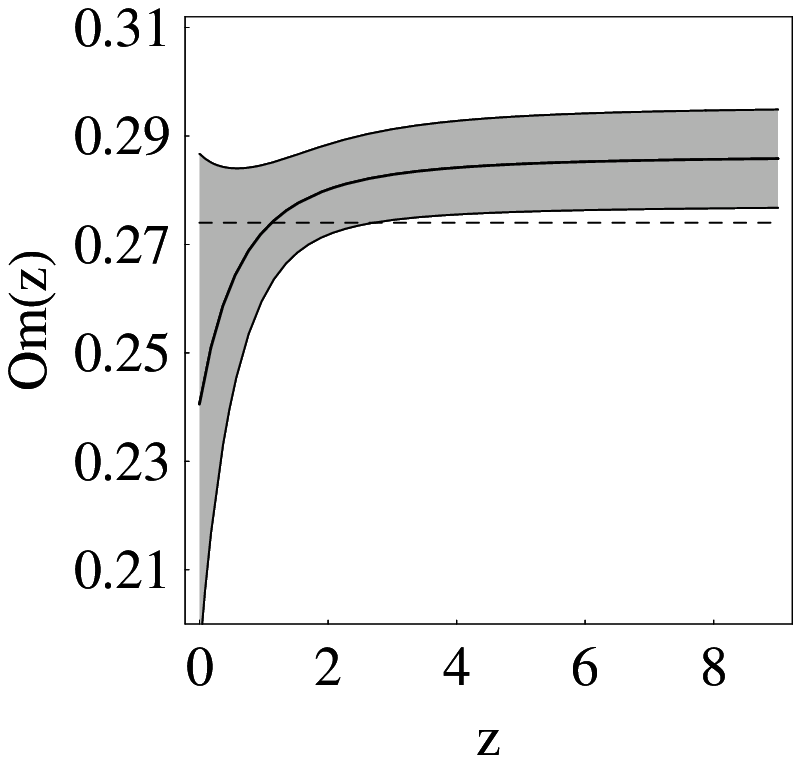}
    ~~ \includegraphics[width=4cm]{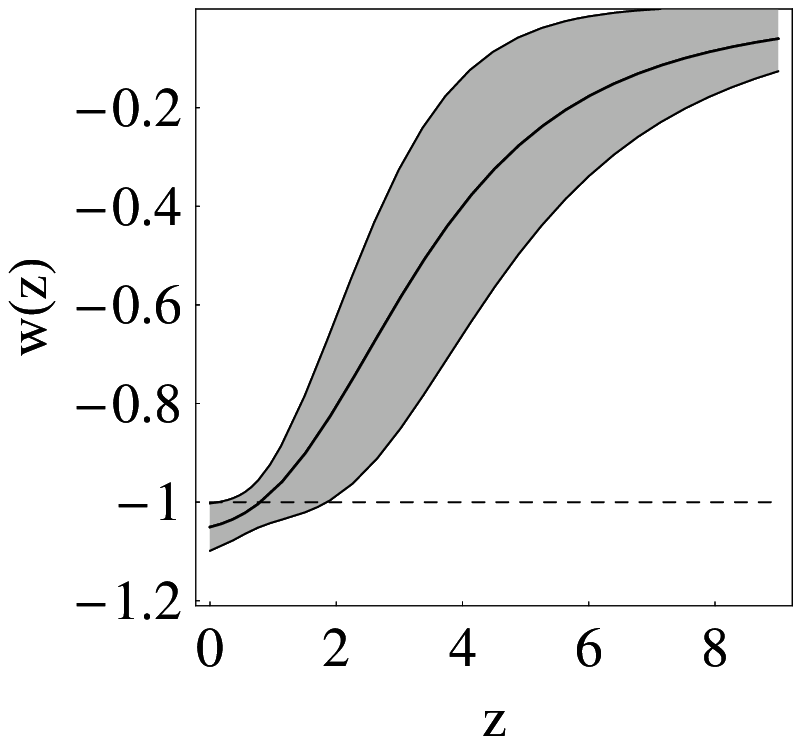}\\
  \caption{The evolution of  $q(z)$,  $Om(z)$ and $w_{de}(z)$  for non-flat generalized holographic model.}\label{figurew-qwz-GH-non-flat}
\end{figure}

\begin{figure}[ht]
  ~~ \includegraphics[width=4cm]{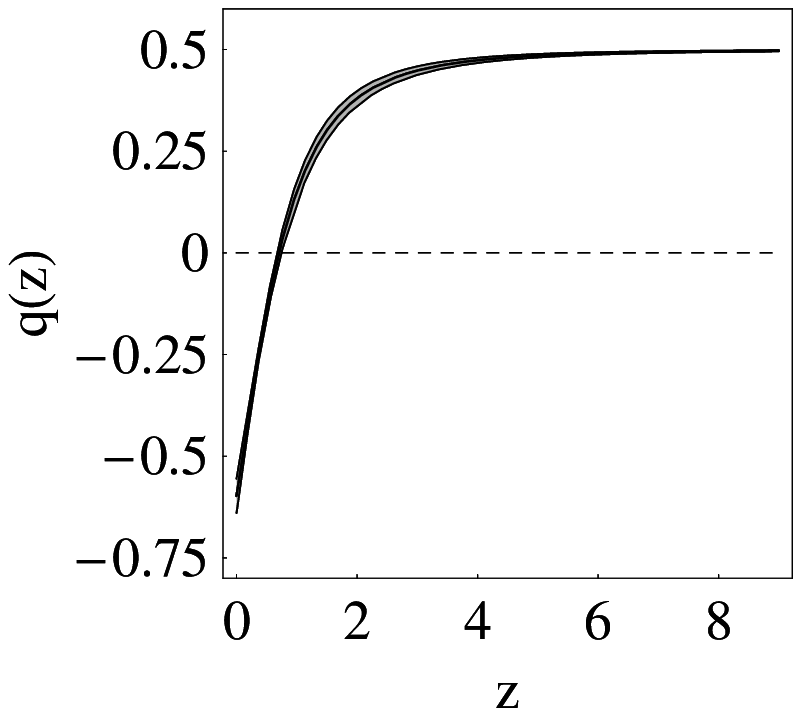}
    ~~ \includegraphics[width=4cm]{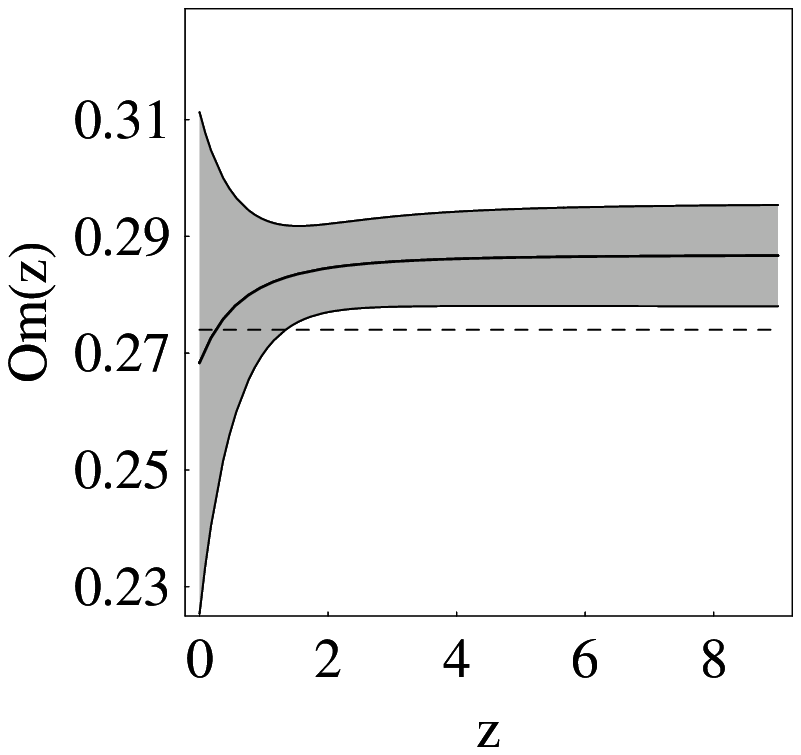}
    ~~ \includegraphics[width=4cm]{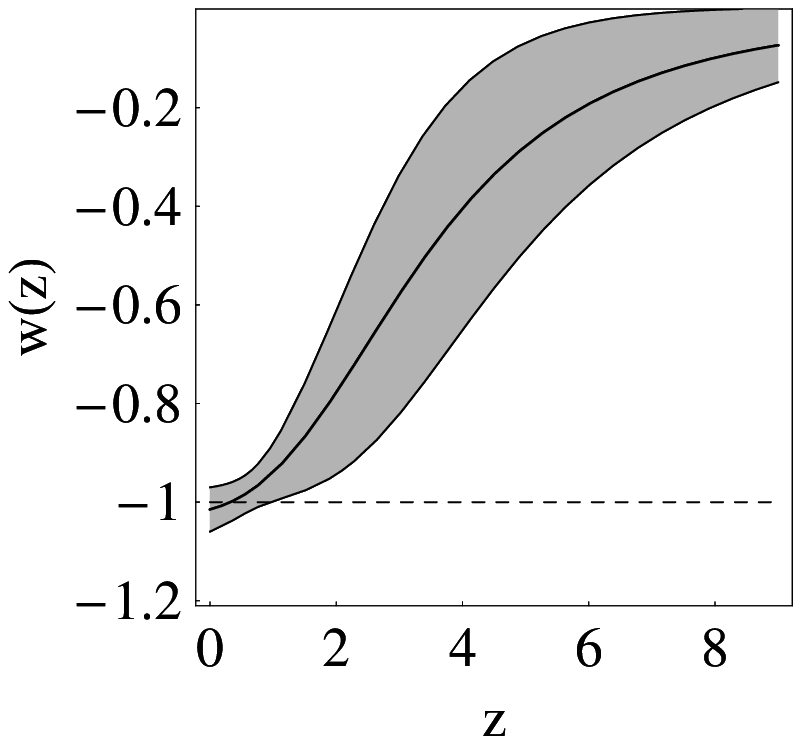}\\
  \caption{The evolution of  $q(z)$,  $Om(z)$ and $w_{de}(z)$  for flat generalized holographic model.}\label{figurew-qwz-GH-flat}
\end{figure}

In addition, according to the calculation of the covariance matrix
and the best fit values of model parameters, the best fit evolutions
of deceleration parameter $q(z)$, geometrical quantity $Om(z)$ and
EOS of dark energy $w_{GH}(z)$ with their confidence level (shadow
region) are plotted in Figs. \ref{figurew-qwz-GH-non-flat} and
\ref{figurew-qwz-GH-flat}. From the figures we can see that a
current accelerated universe is obtained, and the equation of state
for this generalized model can cross over the boundary of
cosmological constant $w_{\Lambda}(z)=-1$. And for this generalized
dark energy model, the predicted values of  some cosmological
parameters with flat and non-flat universe according to above
combined constraints are listed in table \ref{quantities-results}.
From this table, it can be found that the current values of the
deceleration parameter and the EOS of GH model are,
$q_{0}=-0.639^{+0.042}_{-0.047}$, $w_{0GH}=-1.051^{+0.048}_{-0.048}$
for non-flat universe, and  $q_{0}=-0.598^{+0.041}_{-0.042}$,
$w_{0GH}=-1.015^{+0.045}_{-0.045}$ for flat universe. At last as an
appendant, in table \ref{7fgas-results} we also show the best fit
values  of several parameters in $f_{gas}$ analysis method.

By the way, in appendix we also list the constraint  results on
another generalized model in Ref. \cite{GH-xu}, i.e. generalized
Ricci DE by using the MCMC method and above observed data.

\section{$\text{Conclusions}$}
In summary, for interpreting the accelerating universe and solving
the coincidence problems of cosmological constant, the holographic
dark energy models are extensively studied from the different points
of view. In holographic   cosmology, considering that taking the
natural Hubble horizon as the IR cut-off to obtain an accelerated
universe is interesting, Ref. \cite {GH-xu} presents a new
generalized holographic dark energy model.
 In physics, this generalized model investigate a new idea to
 interpret the accelerating universe by using the holographic principle with
 including the Hubble horizon as an IR cut-off. In addition, the
 holographic and Ricci dark energy can be compared in the generalized model
 according to the new introduced parameter $\epsilon$.
In this paper, the flat and the non-flat generalized holographic
dark energy are constrained according to the current observed data.
The stringent constraints on model parameters are given from the
MCMC calculation. Considering the cosmic
 constraint on the parameter  $\epsilon$,
 it is obtained that the cosmic data favor a generalized dark energy
 model which is more Ricci-like, since one has the relation $\rho_{GH}=\epsilon
 \rho_{R}+(1-\epsilon)\rho_{H}$ and the best fit value of
 parameter $\epsilon=1.849$  for a non-flat universe constrained from the observational data.
And according to the constraint results,  it is shown that relative
to the Ricci dark energy model
($\Omega_{0m}=0.300^{+0.037+0.043}_{-0.037-0.042}$ \cite{C-Ricci1}),
it has a smaller value of the dimensionless matter density
$\Omega_{0m}=0.280^{+0.036+0.047}_{-0.032-0.040}$ for the non-flat
universe, which result is more consistent with the current
observations and cosmological constant model \cite{0803.0547}. In
addition, based on the calculation of covariance matrix the best fit
evolutions of cosmological quantities such as
 deceleration parameter, $Om$ parameter and EOS of generalized holographic dark energy
  with their confidence region  are discussed. It is
found that the EOS for this dark energy model can cross over the
boundary of cosmological constant ($w_{\Lambda}=-1$). And the
  values of transition redshift, current deceleration parameter, EOS of GH dark energy are
  obtained, respectively. It can be seen that for the flat universe the best fit value
  of $w_{0GH}=-1.015^{+0.045}_{-0.045}$ is near to the cosmological
  constant model.

 \textbf{\ Acknowledgments }
 The research work is supported by the National Natural Science Foundation
  (Grant No. 10875056) and NSF (10703001)  of P.R. China.

\appendix
\section{$\text{cosmological combined constraints on generalized Ricci DE model by using MCMC method}$}

Considering Ref. \cite{GH-xu}, another extended form dubbed as
generalized Ricci dark energy is  expressed,
\begin{equation}
\rho_{GH}=3c^{2}M_{p}^{2}[1-\eta(1-\frac{H^{2}}{R})]R,
\end{equation}
where  $\eta$ is a parameter. It is easy to see when $\eta=1$ or
$\eta=0$, this generalized form  reduces to Ricci or holographic
dark energy, respectively. And  the Friedmann equation is described
for this generalized model as,
\begin{equation}
H^{2}=H_{0}^{2}[(1-\frac{2(\Omega_{0m}+\Omega_{r}+\Omega_{k})}{2-c^{2}(1+\eta)})(1+z)^{\frac{2}{c^{2}(\eta-1)}
+\frac{2(\eta-2)}{\eta-1}}+\frac{2(\Omega_{0m}(1+z)^{3}+\Omega_{r}(1+z)^{4}+\Omega_{k}(1+z)^{2})}{2-c^{2}(1+\eta)}].
\end{equation}
From above equations one can see that two generalized dark energy
models are equivalent when $\epsilon=1-\eta$. Figs.
\ref{figure-ab-GR-mcmc-6data-non-flat}  shows the $1D$ distributions
of model parameters. And the MCMC calculation results for the
non-flat universe are,
 $\Omega_{k}=-0.0008^{+0.0089+0.0128}_{-0.0127-0.0143}$, $c=0.585^{+0.024+0.029}_{-0.045-0.049}$,
 $\eta=-0.855^{+0.374+0.470}_{-0.366-0.448}$, and
 $\Omega_{0m}=0.280^{+0.037+0.050}_{-0.032-0.036}$ for the best fit values.


\begin{figure}[ht]
  \includegraphics[width=14cm]{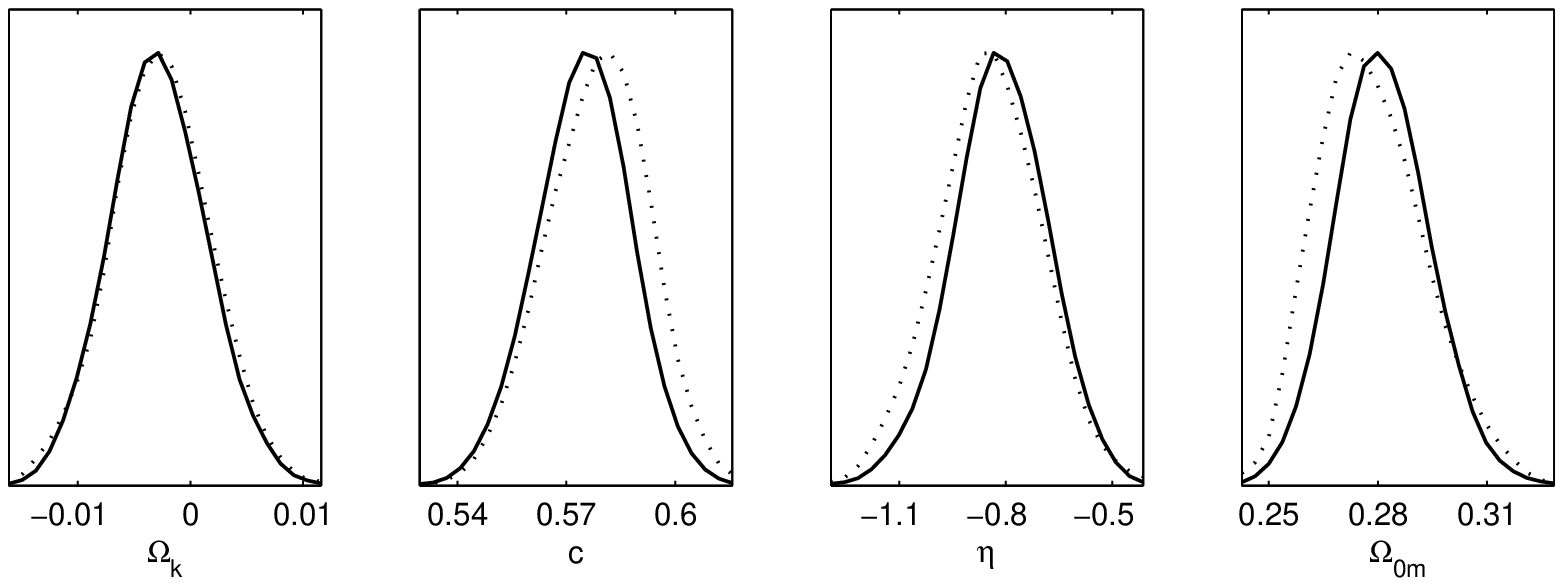}
  \caption{The 1-D
  distribution of model parameters  in the non-flat GR model.
  Solid lines are mean likelihoods
  of samples, and dotted  lines are marginalized probabilities for 1D
  distribution.}\label{figure-ab-GR-mcmc-6data-non-flat}
\end{figure}


\end{document}